\documentclass{article}

\usepackage[nonatbib, final]{neurips_distshift_2022}

\usepackage[utf8]{inputenc} 
\usepackage[T1]{fontenc}    
\usepackage{hyperref}       
\usepackage{url}            
\usepackage{booktabs}       
\usepackage{amsfonts}       
\usepackage{nicefrac}       
\usepackage{microtype}      
\usepackage{xcolor}         

\usepackage{algorithm}
\usepackage{algorithmic}

\usepackage{hyperref}       
\usepackage{url}            
\usepackage{booktabs}       
\usepackage{amsfonts}       
\usepackage{nicefrac}       
\usepackage{microtype}      
\usepackage{xcolor}         
\usepackage{amsmath}
\usepackage{amssymb}
\usepackage{mathtools}
\usepackage{amsthm}
\usepackage[utf8]{inputenc} 
\usepackage[T1]{fontenc}    

\usepackage{color}
\usepackage{caption}
\usepackage{subfigure}
\usepackage{graphicx}
\usepackage{multirow}
\usepackage{wrapfig}
\usepackage{url}

\usepackage[normalem]{ulem}

\def\clap#1{\hbox to 0pt{\hss #1\hss}}%
\def\initials#1{\protect\clap{\smash{\raisebox{1.4ex}{\tiny{\textsf{\textit{#1}}}}}}}%

\makeatletter
\newcommand{\NOTE}[3]{\protect\@ifundefined{hidecomments}{%
  \strut{\color{#2}{\hspace{0pt}\initials{#1}\protect{[#3]}}}%
  }{}}

\title{DSLOB: A Synthetic Limit Order Book Dataset for Benchmarking Forecasting Algorithms under Distributional Shift}

\author{\textbf{Defu Cao\textsuperscript{\rm 1}, Yousef El-Laham\textsuperscript{\rm 2}, Loc Trinh\textsuperscript{\rm 1}, Svitlana Vyetrenko\textsuperscript{\rm 2}, Yan Liu\textsuperscript{\rm 1}}\\
\textsuperscript{\rm 1}University of Southern California~~ 
\textsuperscript{\rm 2} J.P.Morgan AI Research \\
\small \{defucao, loctrinh, yanliu.cs\}@usc.edu ~~ \{yousef.el-laham, svitlana.vyetrenko\}@jpmchase.com
}

\begin{document}

\maketitle

\begin{abstract}

In electronic trading markets, limit order books (LOBs) provide information about pending buy/sell orders at various price levels for a given security. Recently, there has been a growing interest in using LOB data for resolving downstream machine learning tasks (e.g., forecasting). However, dealing with out-of-distribution (OOD) LOB data is challenging since distributional shifts are unlabeled in current publicly available LOB datasets. Therefore, it is critical to build a synthetic LOB dataset with labeled OOD samples serving as a testbed for developing models that generalize well to unseen scenarios. In this work, we utilize a multi-agent market simulator to build a synthetic LOB dataset, named DSLOB, with and without market stress scenarios, which allows for the design of controlled  distributional shift benchmarking. Using the proposed synthetic dataset, we provide a holistic analysis on the forecasting performance of three different state-of-the-art forecasting methods. Our results reflect the need for increased researcher efforts to develop algorithms with robustness to distributional shifts in high-frequency time series data. 

\end{abstract}

\section{Introduction}

Increasingly large market volumes are traded today electronically across multiple asset classes.  Electronic trading is typically facilitated by limit order books (LOBs) - which present the list of orders that is maintained by a trading venue to indicate the “buy” and “sell” interest of market participants for a given security. Specifically, LOBs dynamically record volume and price information  about the buy and sell orders that are being placed in the market at different times \cite{Bouchaud2002}. LOBs present highly complex and noisy environments which enable multiple market participants to trade.
 In addition, trading is usually performed using automated trading algorithms - part of which assumes the ability to forecast LOB prices and volumes \cite{Almgren1999, Sirignano2016}. 

Distributional shift refers to the fundamental issue that the underlying distributions of training and testing datasets are different from each other, which often causes machine learning systems to fail in handling out-of-distribution (OOD) inputs \cite{distrshift2020, generalizations2020}. For example, financial downstream tasks involving time series data can suffer from OOD inputs over time as a result of exogenous factors  (e.g.,  macro shocks, earnings announcements, global pandemic, etc.). Moreover, such sudden distributional  shifts can affect the LOB trading algorithms, specifically their price and volume prediction components.

Different from  the computer vision (CV) and natural language processing (NLP) downstream tasks, where one can easily confirm whether the data suffers from distributional shift ~\cite{ng2020ssmba, hendrycks2018deep, NEURIPS2019_1e795968, liu2020energy}, the correlation structure of multivariate time series data is inherently different than that of images and text. Therefore, previous distributional shift algorithms applied to the CV and NLP domains are not necessarily suitable for the time series domain. Furthermore, publicly available datasets typically used for benchmarking forecasting algorithms are difficult to be used directly to verify whether a particular model has the ability to handle OOD~\cite{cao2021spectral, zhang2022counterfactual}. There are two major reasons: (1) free access to publicly available LOB data is limited, and (2) distributional shifts (e.g., market shocks) are not labeled in real data, making the assessment of a model's ability to account for distributional shifts difficult.

Typically, test performance on OOD inputs is worse than that of test inputs following the same distribution, i.e., independent and identically distributed (IID), as the training data. The over-reliance on IID inputs makes machine learning systems challenging to deploy in real-world settings, where distributional shifts are common ~\cite{yang2021generalized, ramponi2020-neural}. However, to the best of our knowledge, distributional shifts on financial market data have not been thoroughly explored, despite the fact that they are common in real-world financial markets. In this work, we aim to consider a domain adaption task on LOBs, where training and testing data are generated from related but different domains via a multi-agent market simulator.  To this end, we propose a synthetic dataset named DSLOB, which can be used to benchmark forecasting algorithms on a variety of different downstream tasks, e.g., mid-price trend prediction under distributional shifts. Our proposed dataset can test the OOD adaption capabilities of state-of-the-art forecasting models, which should be expected to generalize to unseen samples in spite of the occurrence of a distributional shift. Specifically, we utilize the multi-agent limit order book market simulator called ABIDES~\cite{byrd2019abides} to build a synthetic LOB dataset, where randomly introduced shock are utilized to construct distributional market shifts (i.e., OOD inputs)\cite{wiles2021fine}. Since the LOB dataset is generated in a controlled manner, each snapshot of the LOB is labeled, allowing for straightforward benchmarking on IID vs. OOD inputs. Furthermore, our proposed configuration of the simulator allows for the parametric characterization of multiple types of shocks (e.g., shocks of different magnitudes), which can be used to understand the robustness of each forecasting algorithm as a function of shock parameters.

\paragraph{Contributions:} We summarize the main contributions as follows:

\begin{itemize}
    \item  To facilitate research in both machine learning and finance communities, we propose the DSLOB that allows us to model distributional shifts due to market shocks in LOBs data using a bottom-up multi-agent approach. This approach allows for easy adjustment of the parameters of the market agents to model a wide spectrum of counterfactual shock scenarios. 
    \item To apply a rigorous comparison on the proposed synthetic dataset, we choose three categories of time series forecasting algorithms including (1) AdaRNN-based method focusing on distributional shift problem of time series data; (2) transformer-based method dealing with traditional time series prediction task with long term dependencies; and (3) DeepLOB specializing in high-frequency LOB data.
    
    \item Evaluation results on both IID setting and OOD setting support two primary conclusions: (1) domain shifts can cause algorithms without considering OOD generalization fail to work; (2) there is significant room for improvement in machine learning solutions focusing on time series with OOD samples.
   
\end{itemize}

\section{Limit Order Book Data}

\begin{wrapfigure}{l}{0.450\textwidth}
\setlength{\abovecaptionskip}{0cm}   
    \centering
    \includegraphics[width = 0.45\textwidth]{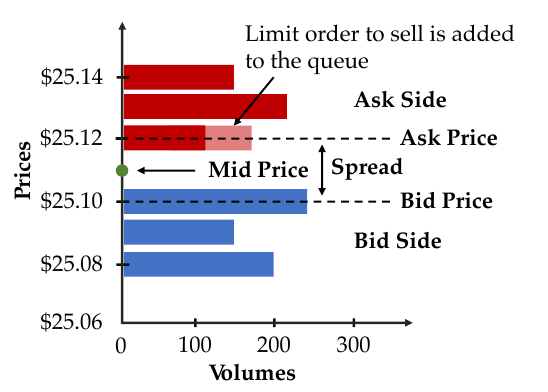}
    \caption{A example of LOB data }
    \label{lob_data_main}
\end{wrapfigure}

As shown in Figure.~\ref{lob_data_main},  a LOB record represents a snapshot of the supply and demand for a given security at a given time instance. It serves as a record of all the outstanding buy (ask)/sell (bid) orders organized by price levels. Additionally, LOBs provide information about the order size (volume) at each price level.  Mathematically speaking, one can regard a snapshot of a LOB as a matrix, where each row vector corresponds to the associated price and order size of the traded asset at a particular level. In LOBs, order types can either be limit orders or market orders. A limit order specifies a price level at which a trader is willing to buy or sell the asset of interest. In other words, limit orders are passive orders in the LOB on the side of the book of the market participant (buy or sell side). In contrast, when a trader places a market order, it indicates that they are willing to buy or sell the asset at the best available price. 
In this work, we use 100 most recent LOB records with 40 features as the input $X = [x_1, x_2,\dots , x_{100}]^T$, where $x_t = [p_a^i(t), v_a^i(t), p_b^i(t), v_b^i(t)]_{i=1}^{n=10} $. In addition, for the given LOB data at time $t$, $p^i$ denotes the price  and $v_i$ denotes the volume size at $i$-th level on both ask side ($p^i_a$) and bid side ($p^i_b$). Therefore, the mid-price $p_t= \frac{p^1_a(t) + p^1_b(t)}{2}$ can be used to create the label $y_t = p_t$ for price forecasting.

\subsection{Generation of LOB Data Using Multi-Agent Market Simulation}

The synthetic dataset used for benchmarking purposes was generated using a multi-agent limit order book market simulator called ABIDES~\cite{byrd2019abides}. ABIDES is an event-based simulation environment that is composed of a simulation kernel, a single exchange, and various market participants. The simulation kernel manages the flow of time and handles all inter-agent communication. For example, all requests/orders completed by background agents (e.g., placing limit and market orders) are managed via the simulation kernel. The exchange is a NASDAQ-like exchange agent that lists security for trade against a LOB with FIFO matching rules. The market participants are so-called background agents that represent market agents with different types of trading strategies. 

\subsubsection{Background Agents Descriptions for ABIDES}

In this work, we mainly looked at the following types of background agents:
\begin{itemize}
    \item {\bf Noise agents} are non-strategic agents that do not base their trading actions on intelligent strategies. In particular, noise agents place limit orders of random size (volume) and of random direction (buy or sell) with interarrival times independently sampled from a discrete uniform distribution from 1 to 100 nanoseconds. 
     \item {\bf Momentum agents} are agents that place market orders based on mid-price trends. More precisely, given lookback periods $T_{\rm min}$ and $T_{\rm max}$ where $T_{\rm max}>T_{\rm min}$, momentum agents use a moving average filter to compute  the average mid-price over each of the lookback periods. If the mid-price based on the shorter lookback period is larger than that of the longer lookback period, intuitively, the momentum agent believes the price is increasing and will place a buy order of random size (or sell order in the opposite case). Unlike value agents, momentum agents are configured to have deterministic interarrival times, i.e., they arrive at the market every $T_{\rm MOM}$ seconds.
     \item {\bf Market maker agents} are agents that supply liquidity to the market by placing orders on both sides of the LOB at various price levels every $T_{\rm MM}$ seconds. For more information about the market maker used in this configuration of agents, please see \cite{vyetrenko2020get}.
     \item {\bf Value agents} are strategic agents that base their trading actions based on an internal estimate of the fundamental value of the asset being traded obtained from some noisy observation. If the estimate of the fundamental value implies that the price of the asset will go up, then the agent will place a buy order. On the other hand, if the noisy observation of the fundamental implies that the price of the asset will go down, then the agent will place a sell order. Let $x_t$ denote the fundamental value at time $t$ and $y_t$ denote its corresponding noisy observation. In ABIDES, the fundamental is modeled via an Ornstein-Uhlenbeck (OU) process. An OU process is a mean-reverting process and the probability density function fundamental value at time $t'$ given the fundamental value at time $t<t'$ is Gaussian, i.e.,
    \begin{equation}
        \label{eq: fundamental_ou}
        p(x_{t'}|x_{t}) = \mathcal{N}\left(x_{t'}|\mu + (x_t-\mu)e^{-\theta \Delta_t}, \frac{\sigma_x^2}{2\theta}(1-e^{-2\theta\Delta_t})\right),
    \end{equation}
     where $\Delta_t = (t'-t)$, $\mu$ is mean of the process, $\sigma_x$ is the volatility of the process, and $\theta$ is the mean-reversion parameter. The value agent's observation model of the fundamental is also Gaussian, i.e., at time $t'$ the value agent believes that the observed fundamental $y_{t'}$ is simply a Gaussian perturbation of the true fundamental: 
     \begin{equation}
         \label{eq: observation_model}
         p(y_{t'}|x_{t'}) = \mathcal{N}(y_{t'}|x_{t'}, \sigma_y^2), 
     \end{equation}
     where $\sigma_y^2$ is the observation noise variance. Given the parameters of the OU process and the observation model, value agents obtain an estimate of the fundamental using a simple Bayesian estimation procedure to determine an estimate of the true fundamental  that is used to drive their trading decision.
     The arrivals of a value agent are modeled as a Poisson process in a standard configuration with mean arrival rate $\lambda_{\rm value}$.
\end{itemize}
Given a configuration which specifies the number of each type of agent, as well as their corresponding parameters, ABIDES can be used to simulate LOB data. Next, we will describe  how distributional shift was modeled into the LOB dataset.

\subsubsection{ABIDES with Distributional Shifts}
To introduce a distributional shift to LOB data, we modify the trading behavior of a type of background agent called a value agent. Value agents are strategic agents that base their trading actions on an internal estimate of the fundamental value of the asset of interest. Value agents derive these estimates from noisy observations of the true fundamental (e.g., a mean-reverting process), which is managed by ABIDES. If the estimate of the fundamental value implies that the price of the asset will go up, then the agent will place a buy order. Similarly, if the estimate of the fundamental value implies that the price of the asset will go down, then the agent will place a sell order. We make two important changes to the design of value agents in order to introduce distributional shifts into our market data:
\begin{enumerate}
    \item We introduce a Gaussian shock $S$ to the observed fundamental that occurs at random time $T_s$ in a random direction $d_s\in\{-1, 1\}$, i.e., 
    \begin{equation}
        \label{eq: shock_distribution}
        S \sim \mathcal{N}(d_s\mu_s, \sigma_s^2), 
    \end{equation}
    where $\mu_s$ and $\sigma_s^2$ denote the mean and variance of $S$, respectively, and $P(d_s=-1)=P(d_s=1)=0.5$. Once the value agents observe the shocked fundamental, their belief about the value of the asset will change, leading to a strong shift in their trading actions.
    \item To emulate agent behavior under realistic shocks, we modify the counting process driving the arrival of value agents. Specifically, rather than using a homogeneous Poisson process to model value agents arrivals, we consider a non-homogeneous Poisson process with arrival rate function $\lambda_{\rm value}(t)$. In the presence of a shock, agents arrive at a higher rate, where in this work, we consider the following arrival rate function:
    \begin{equation}
        \label{eq: arrival_rate_agents}
        \lambda_{\rm value}(t) = \begin{cases}
        \bar\lambda_{\rm value},\quad t < T_s\\
        \bar\lambda_{\rm value}(1+A_s\exp(-\theta_s(t-T_s)), \quad t\geq T_s
        \end{cases},
    \end{equation}
    where $A_s$ is a hyperparameter controlling the scaling factor of the arrival rate and $\theta_s$ controls the reversion of the arrival rate to the mean value $\bar\lambda_{\rm value}$. Intuitively, agent arrivals will arrive with a constant arrival rate $\bar\lambda_{\rm value}$ before the shock occurs. Once the shock occurs, the arrival rate spikes to a value of $(1+A_s)\bar\lambda_{\rm value}$ and decays at exponential rate $\theta_s$ to $\bar\lambda_{\rm value}$ at $t\rightarrow \infty$.
\end{enumerate}
We provide a more detailed configuration in Appendix~\ref{config}, as well as the parameter settings utilized to generate the synthetic LOB dataset used in the experiments of this work.

\section{Deep Learning Benchmarks and Evaluation}

In this work, we can define the distributional shift on LOB data according to the definition of temporal covariate shift~\cite{du2021adarnn}: Given a LOB dataset $D$ with $n$ labeled segments according to the type of shocks, i.e., $D = \{D_i,\dots , D_n\}$, the distributional shift is referred to the case that all the segments under the same type of shock's influence follow the same data distribution $P_{D_i}(x, y)$, while for different types of shocks where $1\leq i \neq j \leq n$ ,  marginal probability distributions are different, and the conditional distributions are the same, i.e.,  $ P_{D_i}(x) \neq P_{D_j}(x)$ and $P_{D_i}(y|x) = P_{D_j}(y|x)$.

We plot the time series mid-price data derived from the generated LOB dataset with a distributional shift in Figure.~\ref{analysis_shock1} and ~\ref{analysis_shcok2} as well as the IID data in Figure.~\ref{no_shock_data} (Refer to Appendix).
After generating this dataset, we adapt three baseline algorithms to the mid-price prediction under the distributional shifts task. Specifically, these baseline algorithms include (1) AdaRNN~\cite{du2021adarnn}, which is a  deep learning model fits for time series distribution-shift problem; (2) Transformer~\cite{zerveas2021transformer}, which is proved the effectiveness of processing time series ~\cite{cao2020spectral, niu2022mu2rest} and (3) DeepLOB~\cite{zhang2019deeplob}, which dedicates to handle LOB data.

\begin{table*}[]
\small
\centering

\begin{tabular}{lccc}

\hline \hline
            & IID & Small Shock & Large Shock \\ \hline
 AdaRNN    & 1.02 $\pm$ 2.24e-4 & 1.00 $\pm$ 8.07e-5 & 0.99  $\pm$3.58e-5 \\ \hline
 Transformer & 0.87 $\pm$ 1.98e-3  & 1.02 $\pm$ 6.29e-3 & 1.08  $\pm$ 0.01 \\ \hline
 DeepLOB     & 0.66 $\pm$ 0.11 & 1.08 $\pm$ 0.07 &	2.25 $\pm$ 0.15  \\ \hline \hline

\end{tabular}
\caption{RMSE results on synthetic LOB dataset. } 
\label{results_label}
\end{table*}

\begin{figure*}[t]
\centering
\includegraphics[width=0.75\linewidth]{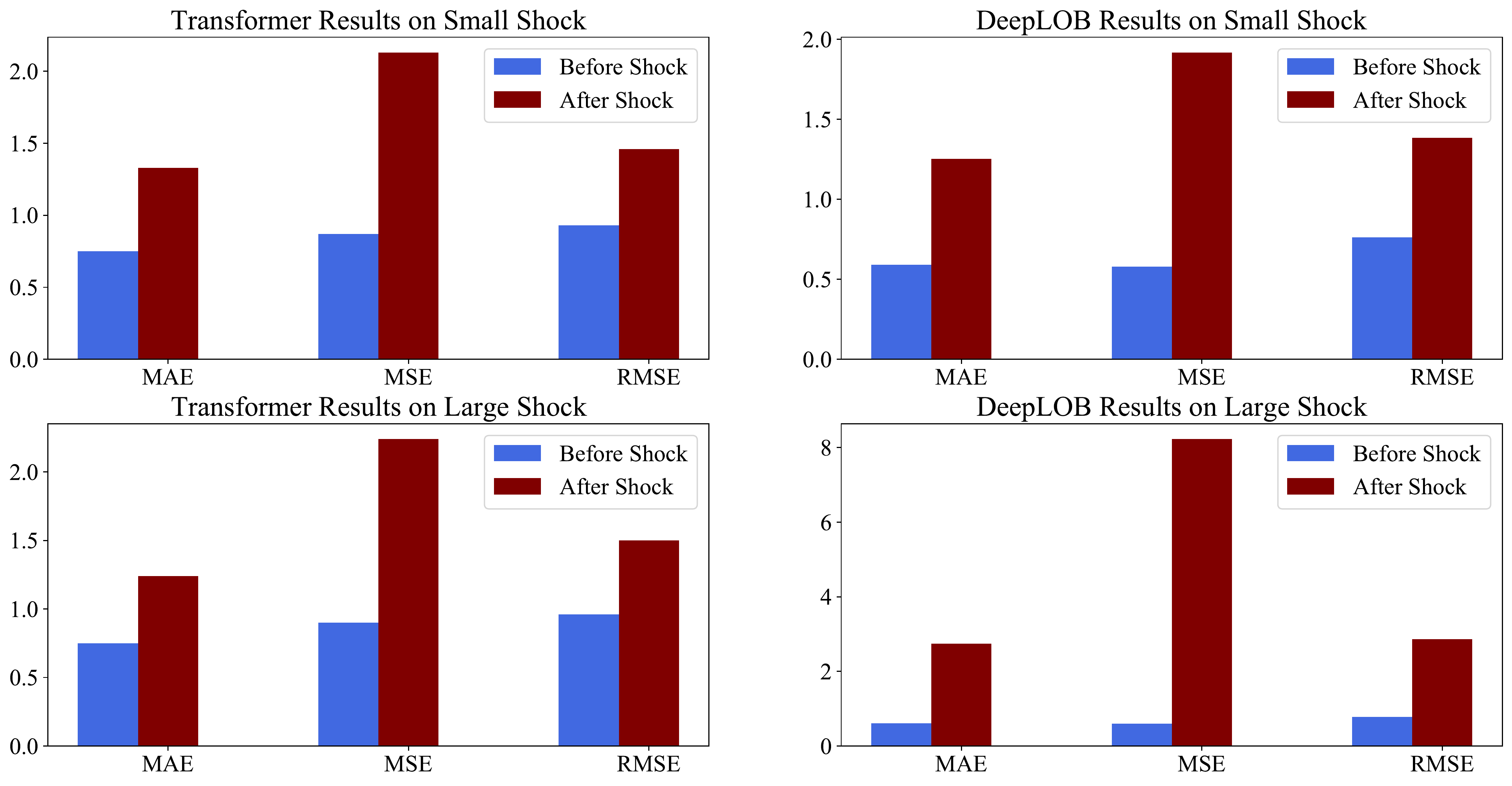} 
\caption{Results on distributional shifts according to the shock type.  }
\vspace{-0.5cm}
\label{fig:shock1}
\end{figure*}

We first use independent and identically distributed (IID) data with no shock coming up, training the baseline models, then we build two settings including the IID set using the test data before the shock appears and the OOD set using the data after the shock appears with the distributional shift. Experimental results are shown in Table~\ref{results_label}, where we can find
 that models without any special treatment on OOD problem (Transformer and DeepLOB) suffer from performance degradation against models dedicated to dealing with distributional shift (AdaRNN).
 In addition,  DeepLOB, which is specifically designed to analyze LOB data, achieves the best performance dealing with IID high-frequency trading data. We further analyze how shocks of different magnitudes impact each model's performance. We plot the performance metrics for Transformer and DeepLOB applied to small and large shocks before and after the shock in Figure.~\ref{fig:shock1}. We find that for a drastic distributional shift (large shock), the performance of the model relying only on the training data degrades more than for the case of a smaller shock. These experimental results suggest that our synthetic dataset can be used to guide the building of machine learning models that are more robust to distributional shifts.

\section{Conclusions}
This work builds a synthetic financial LOB dataset, named DSLOB, which has distributional shifts after some random shocks happen and can be used to verify machine learning benchmarks for financial trend prediction tasks. Next, we evaluate this dataset on three different types of forecasting models and show that considering both distributional shifts and inherent dependencies are still challenging for current machine learning models. In the future, we plan to refine our agent-based model for shock events to cover more granular stress scenarios.

\section{Acknowledgments}
This paper was prepared for informational purposes in part by the Artificial Intelligence Research group of JPMorgan Chase \& Co and its affiliates (“J.P. Morgan”), and is not a product of the Research Department of J.P. Morgan.  J.P. Morgan makes no representation and warranty whatsoever and disclaims all liability, for the completeness, accuracy or reliability of the information contained herein.  This document is not intended as investment research or investment advice, or a recommendation, offer or solicitation for the purchase or sale of any security, financial instrument, financial product or service, or to be used in any way for evaluating the merits of participating in any transaction, and shall not constitute a solicitation under any jurisdiction or to any person, if such solicitation under such jurisdiction or to such person would be unlawful.  This material was partially supported by the J.P. Morgan Faculty Research Award.

\bibliography{camera_ready}
\bibliographystyle{plain}

\newpage
\appendix

\section{Related Work on Financial Datasets.}

There are no unifying benchmark datasets for financial tasks. For instance, S\&P 500~\cite{denis2003s}: Standard \& Poor’s 500 Index is a market-capitalization-weighted index of the 500 largest U.S. publicly traded companies. NASDAQ index~\cite{elliott2003price}: The NASDAQ is the world’s fast electronic stock exchange which operates through computers and telephones, as opposed to traditional methods. NASDAQ lists only technology-based companies. Shanghai stock exchange~\cite{drew2003firm}: The Shanghai Stock Exchange (SSE) represents the largest stock exchange in China. It is a non-profit organization run by the China Securities Regulatory Commission (CSRC). Stocks, funds, bonds, and derivatives are all traded on the exchange. Without the definition of a unique dataset and appropriate performance indicators, researchers cannot make a complete comparison between the proposed studies in order to select a suitable solution for a specific problem. Also, the majority of investigated primary studies provide different evaluation metrics for time-series forecasting. However, most of them  are extracted individually by authors using different splits and date ranges, which may introduce annotators bias. Thus, there is an urgent need for a comprehensive suite of real-world tasks that combine a diverse set of datasets of various sizes coming from financial institutions. Data split as well as evaluation metrics are important so that progress can be measured in a consistent and reproducible way.

 LOB for a given asset are dispersed across several exchanges, creating a fragmentation of liquidity, which 
poses a problem for empirical studies. As ~\cite{Gould2010LimitOB} points out, differences between matching rules and transaction costs of different trading platforms complicate comparisons between different limit order books for the same asset. However, these issues related to fragmentation are not present in the data obtained from the less fragmented Nasdaq Nordic market. In addition, the Helsinki Exchange is a pure limit order market in which market makers have a limited role. For research purposes, the FI-2020 dataset collects high-frequency limit order data of five stocks from the Nasdaq Nordic stock market in 10 consecutive days.  \cite{Balch2019HowTE} show that the Interactive Agent-based Simulation (IABS) market environment can be adapted for using as a backtester. LOB-ITCH \cite{Ntakaris2018BenchmarkDF} includes  5 assets over 10 trading days from Nasdanq Helsinki SE, 4,000,000 observations. The proposed task of LOB-ITCH is  a classification of mid-price movements for 1, 2, 3, 5, and 10 predicted horizons. \cite{Ntakaris2018BenchmarkDF} uses ridge regression and MLP-like network regression methods to verify the quality of LOB-ITCH.

\begin{figure*}[t]

\centering
\subfigure[Day 5 without shock coming]{
\begin{minipage}[t]{0.45\linewidth}
\centering
\includegraphics[width=\textwidth]{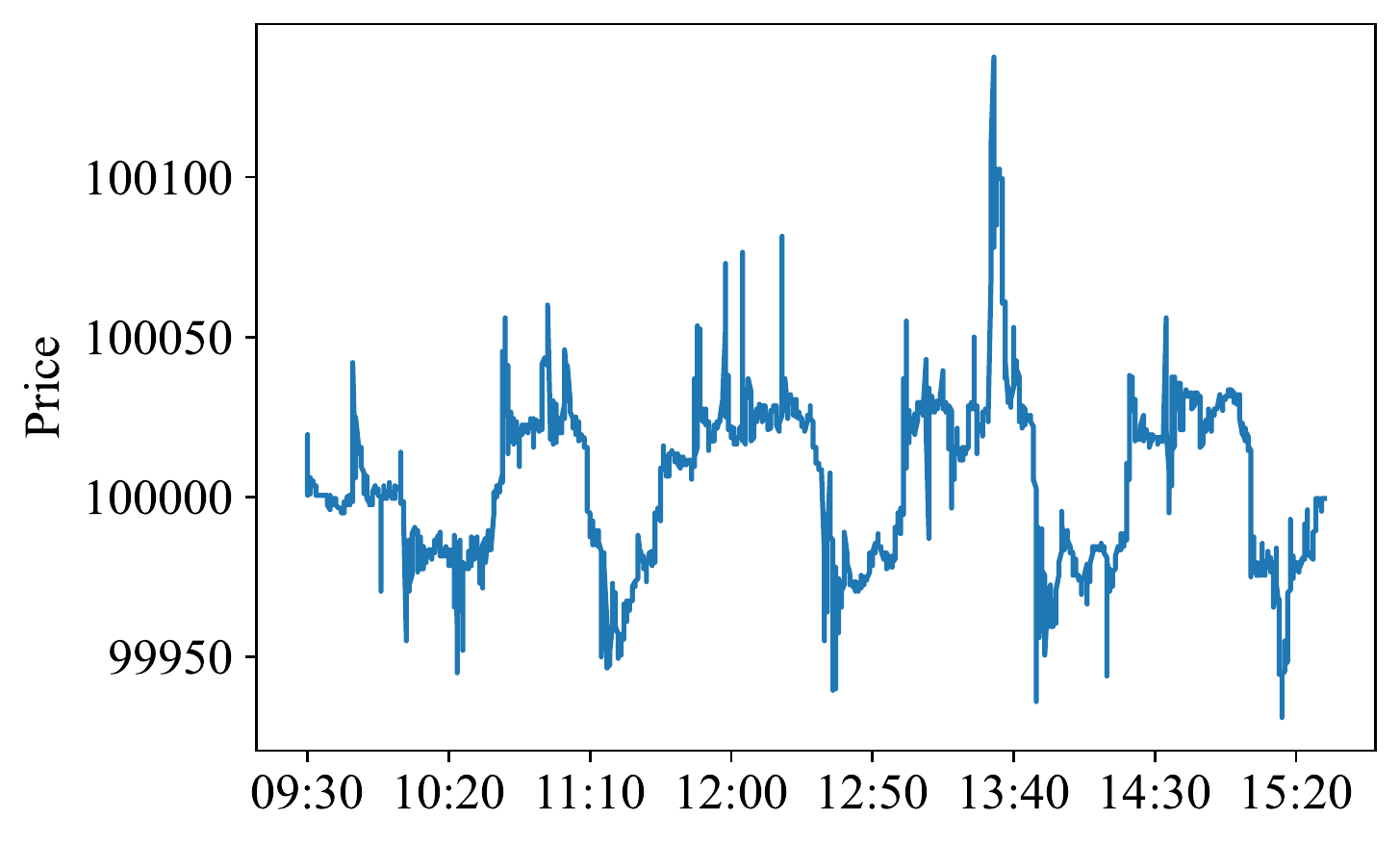}
\label{analysis_fin1}
\end{minipage}%
}%
\subfigure[Day 8 without shock coming]{
\begin{minipage}[t]{0.45\linewidth}
\centering
\includegraphics[width=\textwidth]{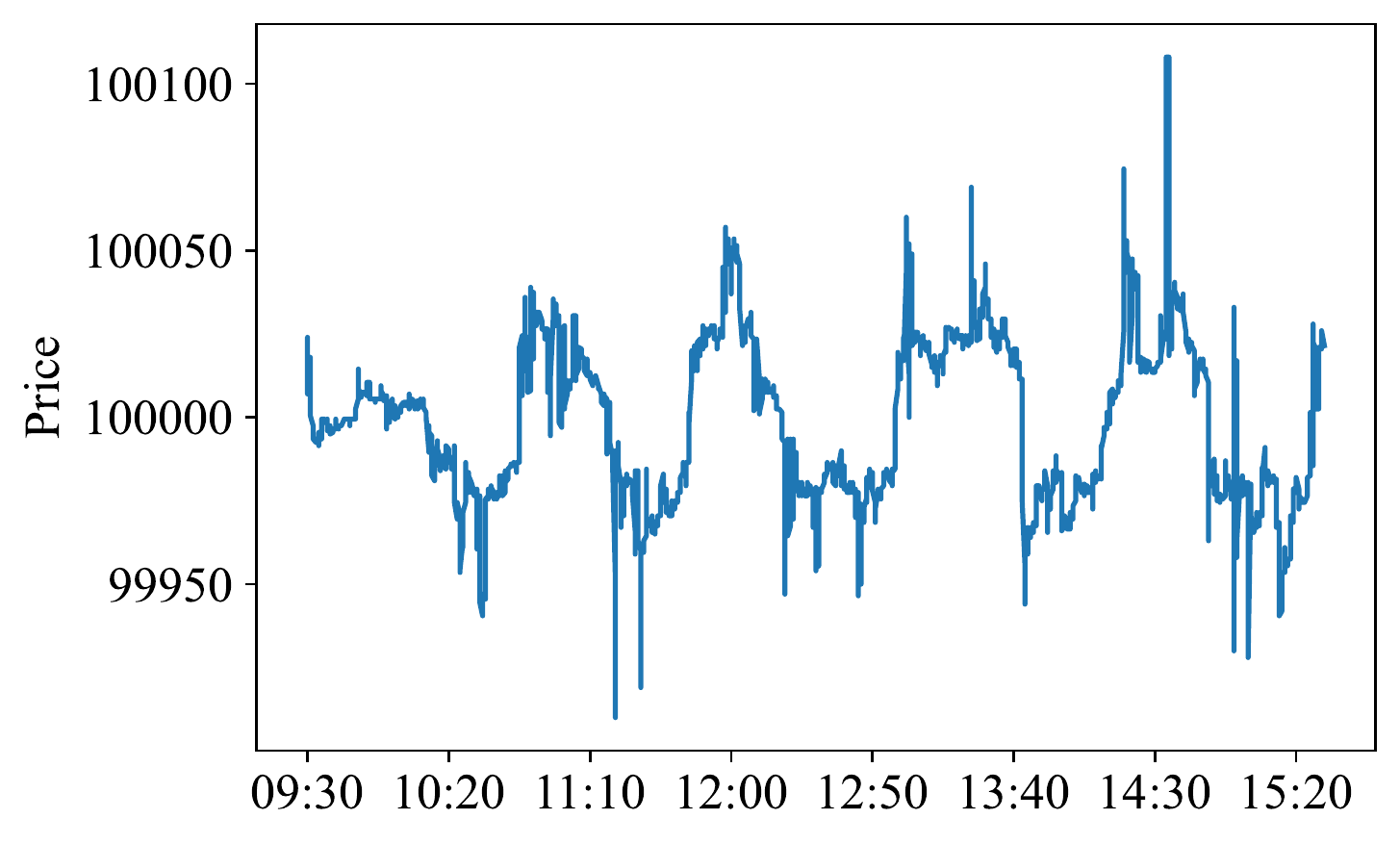}
\label{analysis_fin2}
\end{minipage}%
}%
\caption{Examples of training data's mid-price.}
\label{no_shock_data}
\end{figure*}

\begin{figure*}[t]

\centering
\subfigure[Day 10 with small shock coming]{
\begin{minipage}[t]{0.45\linewidth}
\centering
\includegraphics[width=\textwidth]{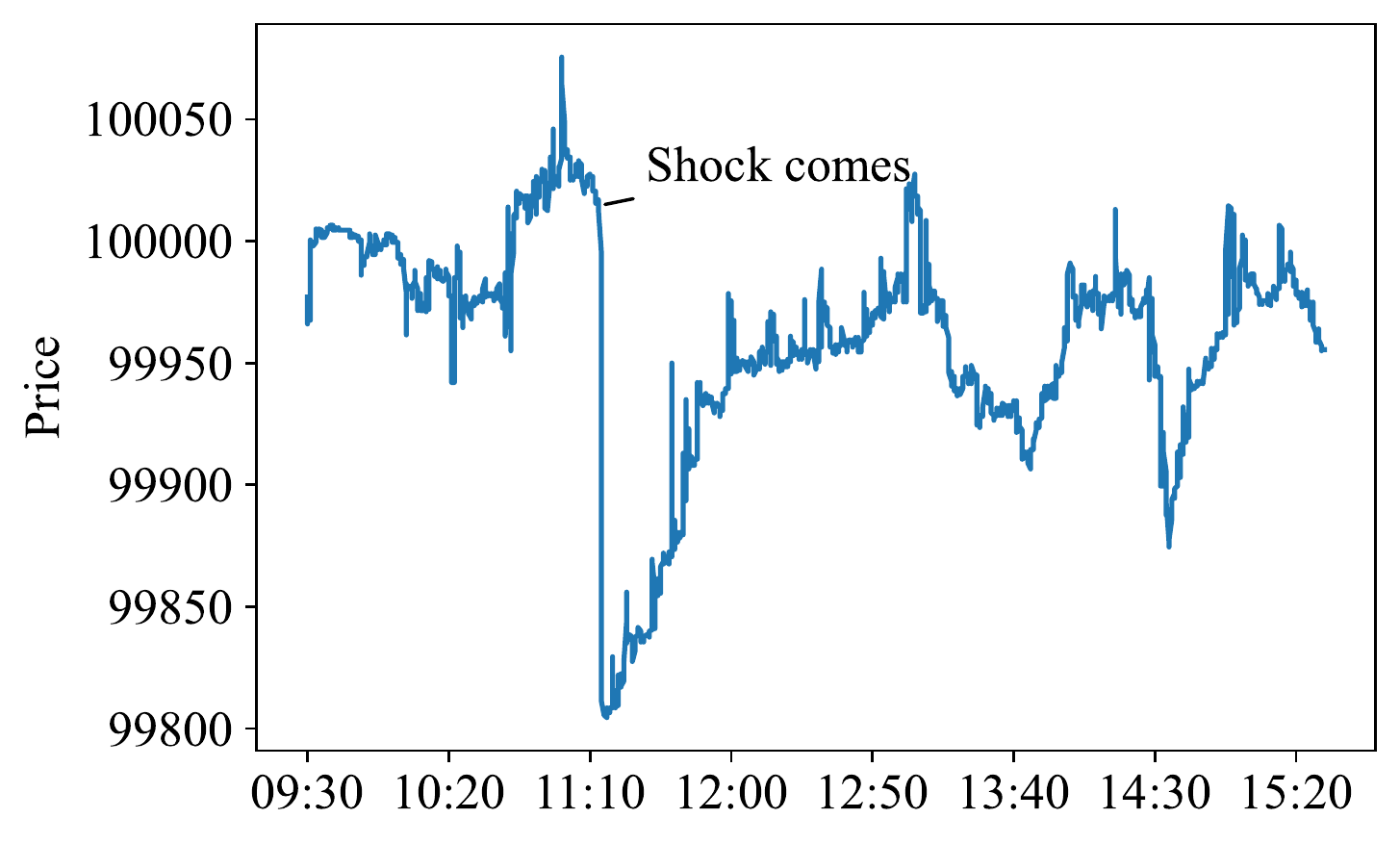}
\label{analysis_fin1}
\end{minipage}%
}%
\subfigure[Day 120 with small shock  coming]{
\begin{minipage}[t]{0.45\linewidth}
\centering
\includegraphics[width=\textwidth]{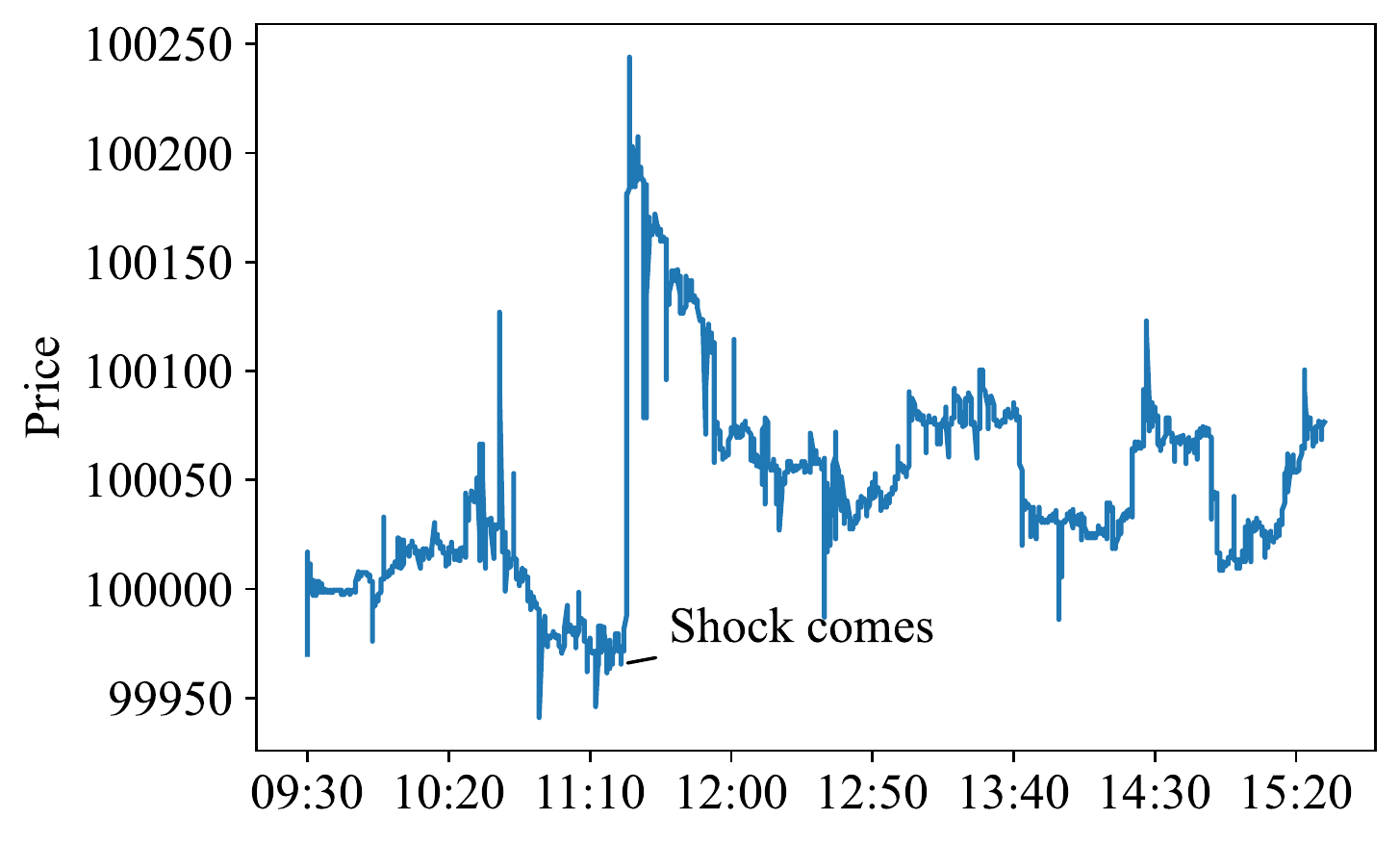}
\label{analysis_bb}
\end{minipage}%
}%
\caption{Examples of distributional shift (small shock) on synthetic data's mid-price.}
\label{analysis_shock1}
\end{figure*}

\begin{figure*}[t]
\centering

\subfigure[Day 2 with large shock coming]{
\begin{minipage}[t]{0.45\linewidth}
\centering
\includegraphics[width=\textwidth]{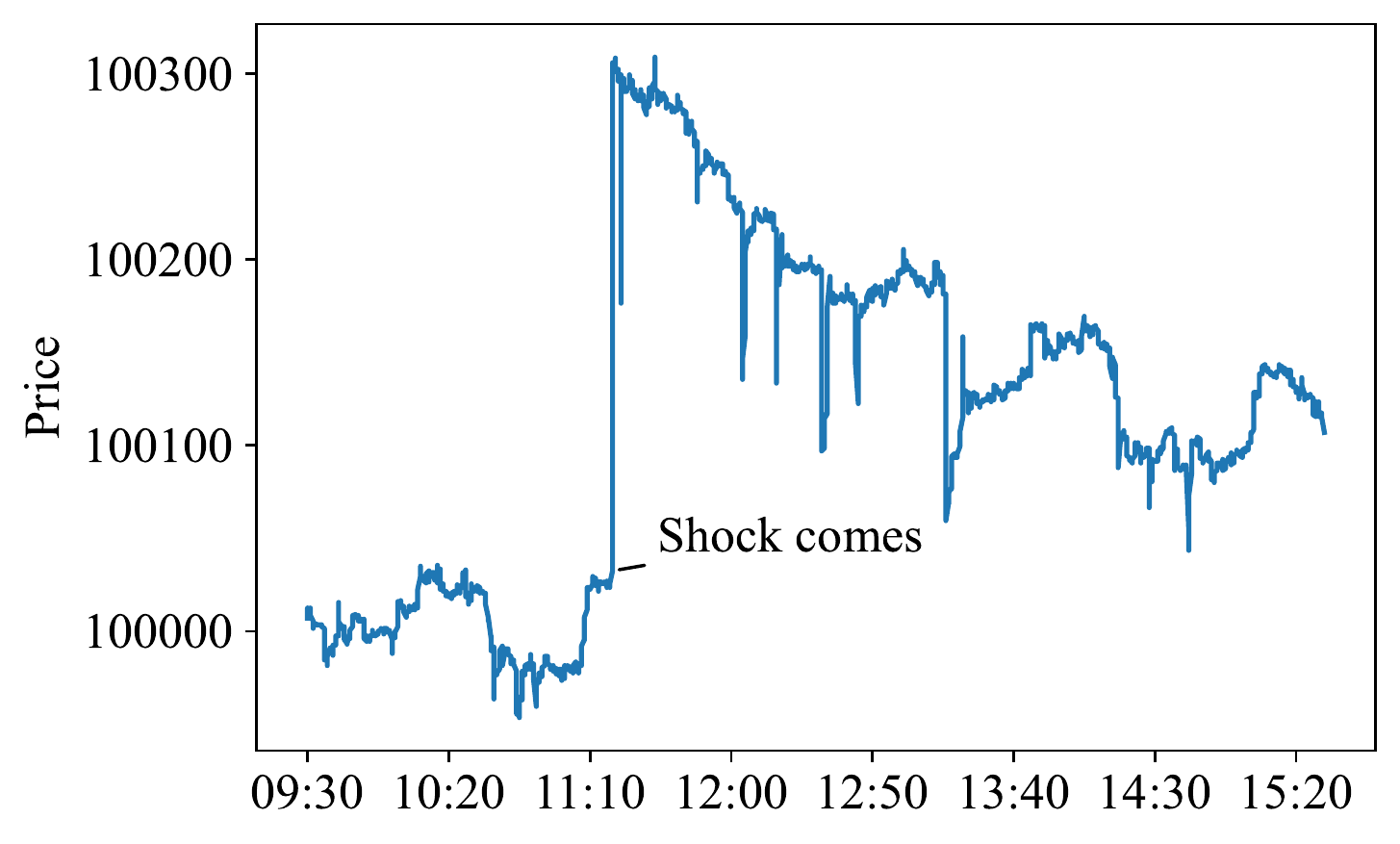}
\label{analysis_fin2}

\end{minipage}%
}
\subfigure[Day 19 with large shock coming]{
\begin{minipage}[t]{0.45\linewidth}
\centering
\includegraphics[width=\textwidth]{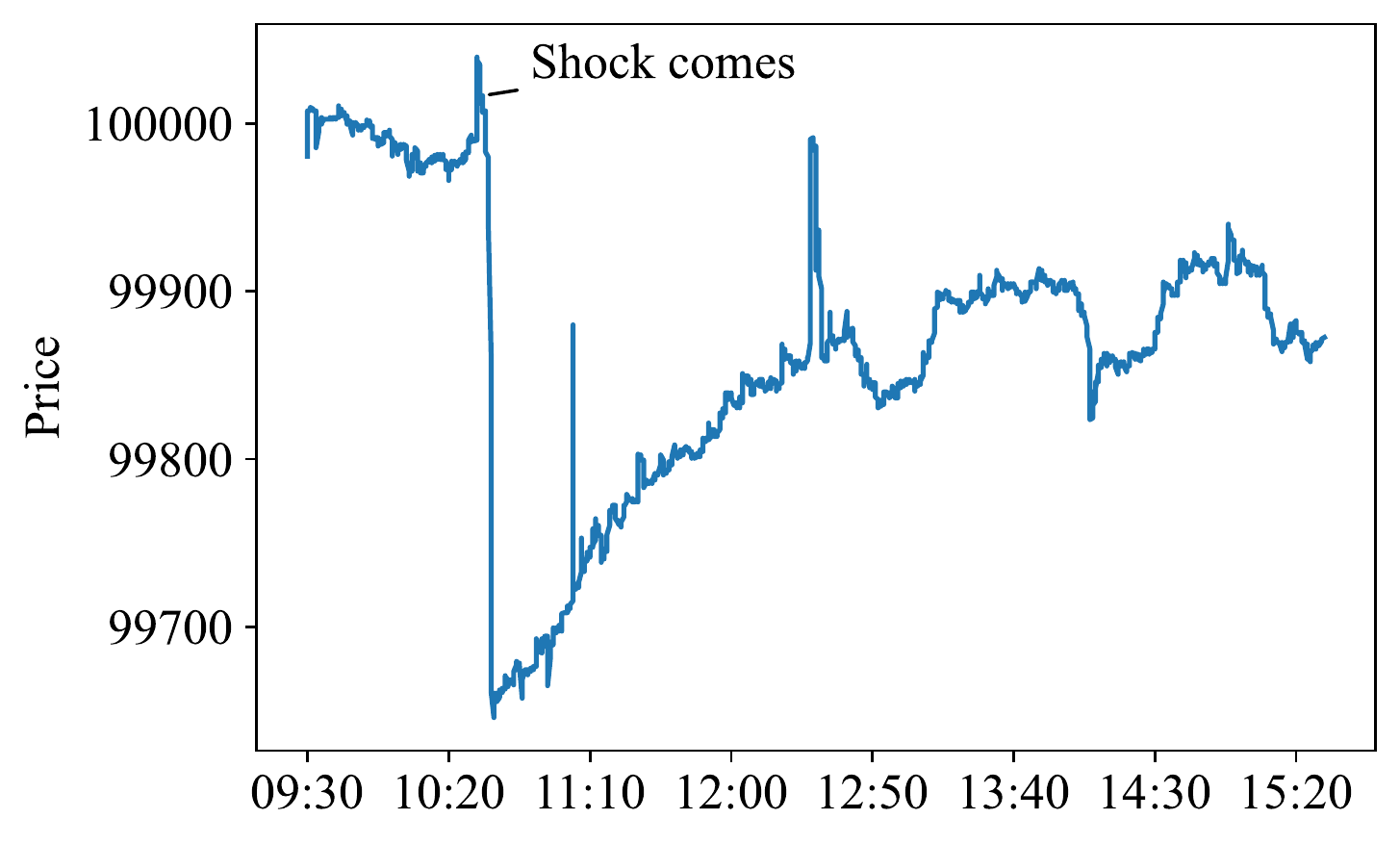}
\label{analysis_fin1}
\end{minipage}%
}%
\caption{Examples of distributional shift (large shock) on synthetic data's mid-price.}
\label{analysis_shcok2}
\end{figure*}

\section{Related work on Distributional Shift Downstream Tasks}

As a way to capture this failure of machine learning models on distributional shifts, many works have been done to find and normalize datasets with out-of-distribution samples~\cite{Koh2021WILDSAB, Ye2021OoDBenchQA}. According to the availability information from the target domain, we can split the distributional shifts tasks into domain adaptation~\cite{wang2018deep, wilson2020multi, hu2020datsing} and domain generalization~\cite{ding2021a, zhou2022domain, hu2022causal}, where the test domains can be visible during the training process of the former, while they are not available during the training of the latter.
 Previous work uses deep auto-regressive generative models ~\cite{Ren2019LikelihoodRF}  or GANs ~\cite{ziyu2020hhh} to deal with OOD samples. Specifically, they propose scoring metrics, such as likelihood estimation, to obtain good OOD detectors. Those models have been shown to be effective in evaluating the likelihood of input data and estimating data distributions. For the time series domain, anomaly detection (AD) setting can be used to  solve OOD problems ~\cite{DBLP2009-11732, DBLP2021lol, zhao2020multivariate}. However, AD  is different from OOD in the following 2 aspects: first, OOD samples cannot be used as labeled examples during the training process in AD  as the distribution  of OOD space is ambiguous; second, AD assumes that the observations of normal samples are homogeneous which will fail to detect OOD samples. For the time series dataset, ~\cite{Zhang2021AnEF} used in-hospital death records and lung x-rays to build distributional shift in the clinical setting, and  ~\cite{guo2022evaluation} used patient health records from the ICU in grouping according to a year of data collection. Besides, ~\cite{Malinin2021ShiftsAD} investigate temporal shifts in a large number of weather data. 
The general framework for working with distributional shift problems was developed in \cite{generalizations2020}.
Several classes of approaches to solving distributional shift problems were tested in \cite{wiles2021fine}.

\section{LOB data}
\label{lob_data_app}

\subsection{Configuration Parameters for Synthetic LOB Dataset}
\label{config}
Using ABIDES, we simulate $N_{\rm days}=365$ trading days worth of data of which 50\% are under ordinary market conditions (no market shocks), and the other 50\% are days that experience a market shock of random magnitude/direction. The configuration includes background agents, of which $N_{\rm noise}=50$ are noise agents, $N_{\rm value}=100$ are value agents with mean arrival rate $\bar\lambda_{\rm value}=0.005$ seconds, $N_{\rm momentum}=10$ are momentum agents with lookback parameters $T_{\rm min}=20$ and $T_{\rm max}=50$, and $N_{\rm MM}=1$ are market makers with wake-up period $T_{\rm MM}=5$ seconds. For all trading days, the parameters of the fundamental are given by: $\mu=100000$ and $\sigma^2=1\times 10^{-12}$.  In total, our synthetic dataset simulates three different scenarios corresponding to both IID and OOD settings:
\begin{itemize}
    \item {\bf Ordinary market conditions:} This comprises 50\% of the simulated data, where no shock is introduced to the fundamental. Under this setting, the mean-reversion parameter of the OU process in this configuration is $\theta=1\times 10^{-12}$. The data produced from this setting are used as the training data for the forecasting algorithms (IID).
    \item {\bf Small shock:} This comprises 25\% of the simulated data and assumes a reversion rate of $\theta=1\times10^{-12}$ on the fundamental. For the shock, we model the shock time according to a uniform distribution $T_s\sim\mathcal{U}(1\ {\rm hours}, 2 \ {\rm hours})$ after the market opens with shock mean and variance as $\mu_s=200$ and $\sigma_s^2=400$, respectively. Furthermore, after the shock occurs, value agent arrivals are governed by the amplification rate $A_s=2$ and arrival rate reversion parameter $\theta_s=1\times 10^{-12}$. The data produced from this scenario are considered OOD inputs. 
    \item {\bf Large shock}: This comprises 25\% of the simulated data and assumes a reversion rate of $\theta=5\times10^{-13}$ on the fundamental. For the shock, we model the shock time according to a uniform distribution $T_s\sim\mathcal{U}(1\ {\rm hours}, 2 \ {\rm hours})$ after the market opens with shock mean and variance as $\mu_s=400$ and $\sigma_s^2=1600$, respectively. Furthermore, after the shock occurs, value agent arrivals are governed by the amplification rate $A_s=3$ and arrival rate reversion parameter $\theta_s=1\times 5^{-13}$.  This scenario corresponds to OOD setting, where the distributional shift is larger than that of the small shock scenario. Since the reversion rate of the fundamental and arrival rate amplification are smaller in this setting, the simulated market shock also lasts longer. 
\end{itemize}

\section{Benchmarks}

AdaRNN is the specific model dealing with the problem of the distributional shift where statistical properties of time series can change over time.  
The first module of AdaRNN, called Temporal Distribution Characterization (TDC), aims to better characterize the distribution information in the time series. The second module is Temporal Distribution Matching (TDM), which uses a boosting-based procedure to learn the hidden representation and to reduce distribution mismatch in time series. By combining TDC and TDM, AdaRNN can utilize the common knowledge by matching the distribution to learn the efficiency representation and then finish the prediction task with that representation.

The transformer is an encoder and decoder structure to process sequence data, which has achieved excellent performance in several time series tasks with the ability to handle long-term dependence. Each encoder/decoder block consists of a multi-head self-attention module and a position-embedding neural network. In addition, each decoder module inserts a cross-attention module between the multi-head self-attention module and the position-embedding neural network. Unlike LSTM or RNN,  the transformer  uses positional encoding to embed positional information for the input instead of  any iterative or convolutional operations. In the task of  mid-price prediction under distributional shifts,  we use a simple encoder model with  an attention layer for learning representations, and the FFN layer is used to predict the final price.

DeepLOB is a deep neural network architecture containing convolutional layers as well as long short-term memory (LSTM) units for predicting future stock price movements in large-scale high-frequency LOB data. DeepLOB contains three modules, where the CNN module with convolutional and pooling layers extracts features automatically to avoid the limitations of hand-crafted features, then the Inception module ~\cite{Szegedy2015GoingDW} helps to infer local interactions over different time scales. After that, the resulting feature maps are passed to an LSTM unit that captures the dynamic temporal behavior. 
In this work, we will use our synthetic data to verify the stability of DeepLOB in the face of OOD problems.

\end{document}